# Slow Evolution of *rag1* and *pomc* Genes in Vertebrates with Large Genomes.

Bianca Sclavi[1]* and John Herrick[2]*

*corresponding authors

1. LBPA, UMR 8113 du CNRS, ENS Cachan, Cachan, France 94235
sclavi@lbpa.ens-cachan.fr

2. jhenryherrick@yahoo.fr

*Abstract*

Growing evidence suggests that many vertebrate lineages are evolving at significantly different rates. As a first approximation of evolutionary rates, we assessed the amount of neutral (dS) and non-neutral (dN) substitutions that have accumulated within and across sister clades since the time of their divergence. We found that in fish, tetraodontiformes (pufferfish) are evolving at faster rates than cypriniformes (fresh water teleosts), while cypriniformes are evolving faster than elasmobranchs (sharks, skates and rays). A similar rate variation was observed in salamanders: plethodontidae were found to evolve at a rate nearly two fold faster than the hydromantes lineage. We discuss possible explanations for this striking variation in substitution rates among different vertebrate lineages that occupy widely diverse habitats and niches.

*Introduction*

Rates of molecular evolution are known to vary significantly across lineages belonging to the same evolutionary group (Lanfear et al. 2010). Nucleotide substitution rates in birds, for example, are higher in the songbird lineage than in chicken (Nam et al. 2010); while in mammals, rates in the murid lineage are higher than in man. The molecular basis for the observed variation in mutation and substitution rates is complex and poorly understood. DNA replication errors, however, are a major source of endogenous mutations, and mutation rates across the genome have recently been found to correlate with DNA replication timing in fungi, invertebrates and mammals (Wolfe et al. 1989; Chen et al. 2010; Weber et al. 2012) (Stamatoyannopoulos et al. 2009; Lang and Murray 2011; Agier and Fischer 2012). In addition, it has been proposed that substitution rates vary as a result of differing DNA repair efficiencies in a lineage specific manner (Britten 1986).

The intricate interplay between DNA replication and DNA repair systems as the cell cycle progresses suggests that growing reliance on error prone DNA repair systems such as Translesion DNA Synthesis (TLS) and Non-homologous End-joining (NHEJ) of DNA double strand breaks might explain the increase in mutation rate as the DNA synthetic phase, or S phase, of the cell cycle advances (Herrick 2011). Other potential and related explanations concern the compartmentalization of the genome into different forms of chromatin (eg. early replicating euchromatin: EC, and late replicating heterochromatin: HC) (Lande-Diner et al. 2009), which vary in DNA content between lineages and differentially rely on DNA repair systems. It remains unknown, however, if these same repair systems can account for differences in mutation/substitution rates between lineages.

In vertebrates, lineage specific mutation rate variation has been associated with several different but interacting life history traits including body size, generation time and metabolic rate (Martin and Palumbi 1993; Bromham 2011). A generation time effect (GT), for example, has been proposed to account for the decrease in mutation rate resulting from DNA replication errors as the primate lineage evolved (Hwang and Green 2004). Low rates of molecular evolution in some acipensiforme lineages have similarly been attributed to a generation time effect on mutation and substitution rates (Krieger and Fuerst 2002). How GT might impact rates of molecular evolution remains unclear, but GT is known to correlate significantly with genome size (C-value) in both plants and animals (Gregory 2001; Hardie and Hebert 2003; Francis et al. 2008).

Low mutation rates are generally acknowledged to be required for the evolution of large genomes. Hinegardner and Rosen first suggested in 1972 that fish with large genomes are evolving more slowly than fish with smaller genomes (Hinegardner and Rosen 1972). An investigation of evolutionary rates in lungfish (C-value 70 pg) likewise revealed that lungfish are evolving up to two fold more slowly than either frogs or mammals (C-value 3 pg) (Lee et al. 2006). Similar observations have been made on salamanders (Kozak et al. 2005). Consistent with observations of low rates of molecular evolution in taxa with large genomes, other studies in plants, fish and animals revealed a genome size effect on extinction rates and species richness (Vinogradov 2004; Knight et al. 2005; Olmo 2006; Kraaijeveld 2010). Together, these observations suggest that variations in mutation/substitution rates influence the mode and tempo of genome size evolution and rates of diversification in different plant and animal lineages.

To further investigate the association between diversification rate and genome size, we measured substitutions at synonymous (dS) and non-synonymous (dN) coding sites in two nuclear genes, *rag1* and *pomc*, from three different vertebrate groups: fish, frogs and salamanders. Within each group, we selected closely related lineages in order to compare the number of substitutions that have occurred since the lineages diverged. Two sister lineages were selected from cypriniformes, the largest freshwater fish clade. Substitution rates were then compared to substitution rates in closely related lineages from tetraodontiformes (pufferfish) and chondrichthyes (skates, rays and sharks). Similar analyses were performed on anurans (hyla and toads) and urodelae (salamanders).

These studies revealed that rates of molecular evolution appear to be strongly conserved between the sister lineages examined here, but vary significantly between distantly related lineages in the same group. In salamanders, however, two closely related lineages, the plethodontidae and the hydromantes, exhibit a more than two-fold variation in evolutionary rates. As expected, these studies also revealed that large genomes tend to be associated with low rates of molecular evolution. The trend is remarkably reproducible among the lineages examined with the exception of cartilaginous fish. In skates, rays and sharks, genome size varies up to ten-fold (1.2 pg to 12 pg), but, as previously reported, substitution rates remain uniform and extremely low across the respective lineages (Martin et al. 1992). These findings contribute to the growing body of evidence that rates of molecular evolution are highly heterogeneous among vertebrates, and support the notion that organisms with large genomes tend to have lower substitution rates and rates of evolution.

*Results*

*Genome size variation in fish, frogs and Salamanders*

Earlier studies in plants, fish and animals revealed an association of genome size with extinction rates and species richness (Vinogradov 2004; Knight et al. 2005; Olmo 2006; Kraaijeveld 2010). The association between genome size and species richness becomes especially apparent in groups with genome sizes larger than 5 pg in amniotes and 14 pg in plants (Knight et al. 2005; Olmo 2006). We therefore examined the number of species as a function of genome size in three related groups: fish, frogs and salamanders. The genome size of each species was obtained from the Animal Genome Size Database (Gregory et al. 2007).

Figure 1 shows that ray-finned fish have an optimal genome size that tends toward smaller genomes between 1 and 2 pg. In contrast, cartilaginous fish and frogs have an optimal genome size between 3 and 5 pg, and salamanders, which are the least speciose of the three groups, tend to have an optimal genome size of 25 to 30 pg. Given that fish are the most species rich group (ray finned fish: ~24000 species, cartilaginous ~810) compared to anurans (~4000) and urodelae (521) these results support the earlier findings that large genome size negatively impacts species richness in different taxonomic groups.

Previous studies have shown that the variation in genome size in teleost fish approximates a lognormal distribution (Hardie and Hebert 2004). The dataset used here is limited to ray-finned and cartilaginous fish. In agreement with the earlier studies, both data sets fit a log normal distribution (Figure 1); combined data sets for fish, however, approximate a power-law distribution (Supplementary Figure 1). In contrast, the distribution in frogs is approximately gaussian, while the urodelae distribution shows two peaks, one between 25 and 30 pg and the second between 40 and 45 pg, both gaussian. In the first peak there is a slightly higher proportion of Ambystomidae (13% vs 9% in the total population) and Salamandriae (45% vs 33%) and a decreased proportion of Plethodonitae (38% vs 47%), which constitutes the majority of the second peak.

A gaussian distribution indicates that the main mechanisms responsible for genome size variation are additive (randomly occurring deletions and amplifications), whereas lognormal distributions indicate multiplicative effects of varying degrees (genome duplication and polyploidization) (Hardie and Hebert 2004). The ancestral vertebrate lineage is believed to have experienced one or two whole genome duplication events. In contrast, teleost fish have undergone an additional duplication event (the 3R hypothesis), which might have contributed to their faster evolutionary rates compared to all other vertebrates (Robinson-rechavi 1998). Hence, genome size variation in the three different groups examined here appears to follow markedly different modes of genome evolution.

*Evolutionary rates of rag1 and POMC in fish, frogs and salamanders*

The groups of species examined here diverged over widely different time scales. The fish lineages, for example, diverged between approximately 600 and 20 million years ago; the frogs diverged about 200 to 60 million years ago; and the salamanders diverged around 25-14 million years ago. Due to the large differences in divergence times in different lineages and to the large differences in evolutionary rates, we decided to use two different genes to measure synonymous and non-synonymous substitution rates, *rag1* and *pomc*.

The former is a relatively slowly evolving gene (*rag1* core: 79 % nucleotide identity between sharks and mammals) that can be used for measuring the rate of synonymous substitutions (Kapitonov and Jurka 2005). Non-synonymous substitutions between closely related lineages, however, are too few in this gene to assess accurately the relative amounts of nucleotide diversity. In contrast, the *pomc* gene is faster evolving: amino acid identity between human and teleost ACTH is 74%–77% for POMCα and 59% for POMCβ sequences (de Souza et al. 2005). This gene was therefore used to measure the rate of non-synonymous substitutions. The rate of synonymous substitutions, however, is too high to be measured within some of the lineages used here, because of their very old divergence times and concomitant saturation effects that can obscure total amounts of nucleotide diversity over the given time scales.

*Low but heterogeneous substitution rates in the rag1 gene in Salamanders*

The *rag1* gene has the advantage that synonymous sequences are not saturated over the evolutionary distances considered here. Phylogenetic analyses were performed and a neighbor joining tree was generated using MEGA5. The number of synonymous and non-synonymous substitutions per site was calculated using the method of Nei and Gojobori (Tamura et al. 2011). Sister lineages were then selected from the phylogenetic trees according to the availability of their C-values in the Animal Genome Size Database. Initially, five lineages of salamander were identified, with C-values ranging from 20 pg (plethodontidae) to 76 pg (hydromantes) (Sessions 2008).

To assess evolutionary rates, we ascertained the period of time since the sister lineages had diverged. Divergence times were obtained from TimeTree (Hedges et al. 2006) or from values reported in the literature (Table 2). Plethodonton and hydromantes, for example, diverged an estimated 14 million years ago (Mya) according to the fossil record. Figure 2A reveals that when divergence times are accounted for, the hydromantes lineage is evolving up to 3X slower than the plethodontidae. When compared to frogs and toads, genetic diversity is substantially lower in both salamander lineages. Accounting for divergence times, however, reveals that the plethodontontidae and anuran lineages considered here are evolving at similar rates.

*Very high rates of diversification in fish with small genomes*

Several earlier studies revealed that teleost fish have very high rates of molecular evolution, while the more ancient lineages of cartilaginous fish have among the lowest evolutionary rates yet identified (Martin et al. 1992; Wang et al. 2009). Using the *rag1* gene, we repeated the above analyses on tetraodontiformes, cypriniformes and elasmobranchii (skates, rays and sharks). Our findings confirm the earlier observations: the *rag1* gene in tetraodontiformes (C value: 0.3 to 0.5 pg) is evolving at very high rates compared to the cypriniformes (C value: 0.5 to 2.7 pg), while the batoidea and etmopteridae (skates and rays and sharks) with some of the largest genomes (1.9 to 12 pg) are evolving at the slowest rates (Figure 2B). Strikingly, the *rag1* gene in tetraodontiformes is evolving at a rate up to 3X faster than in cypriniformes and up to 6X faster than in cartilaginous fish.

Comparing rates of diversification between closely related lineages revealed that substitution rates in the *rag1* gene are highly conserved in the course of evolution. This strong lineage dependent effect is revealed by the similar dS values for lineages in different groups. The tetraodontidae, for example, all have a very similar, high rate of substitution whereas the cartilaginous fish have less variable and much slower rates of substitution independently of genome size. These observations suggest that mutation rates are themselves evolving as lineages split, and might therefore coincide with speciation events. Conversely, speciation events might be driving changes in mutation/substitution rate (see discussion) (Venditti and Pagel 2010).

Our studies also revealed a proportional increase in dN with respect to dS, which is consistent with earlier findings (Stoletzki and Eyre-Walker 2011): dN/dS values in *rag1* are positively correlated and remain largely constant over all lineages examined here (Supplementary Figure 2). This suggests that positive selection has not been an important factor in governing *rag1* substitution rates in these lineages. The proportional increase in dN as dS values increase might indicate an effect of chromosomal location (Chuang and Li 2004) or chromosomal context on substitution rates for both synonymous and non-synonymous rates of mutation (Stoletzki and Eyre-Walker 2011). Together, these observations suggest that mutation/substitution rates are co-evolving with genome size at both synonymous and non-synonymous sites simultaneously.

*C-value and the rate of evolution of the POMC protein*

Earlier studies revealed that protein coding sequences in fish are evolving at a similar rate to frogs, chicken and opossum, while the substitution rate in elephant shark was significantly lower (Wang et al. 2009). These studies, however, were conducted on a large set of genes that do not account for positional effects (271 genes), and consequently represent a genome wide average in substitution rates. We therefore measured the amount of divergence in the proopiomelanocortin (POMC) coding sequences in those species with known C-values. The POMC gene was selected because the protein is conserved across vertebrates and is not directly involved in specifying morphological features that can be affected by external selective forces (Lee et al. 2006; Dores and Baron 2011).

The comparison of the distribution of distances found within each group shows a decreasing mean and variance in genetic distance as a function of increasing genome size (Supplementary Figure 3). The species within each group were then subdivided into different subgroups according to their respective genome sizes. The distances for pairs of species within each subgroup were then plotted as a function of the subgroup's average genome size (Supplementary Figure 4). The data points in this figure are colored according to their divergence times (Hedges et al. 2006). As expected from a molecular clock-like model, the percent difference in amino acids increases with increasing divergence time. In addition, the distance also decreases with increasing genome size. This indicates that, in addition to lineage specific effects, genome size influences the amount of diversity within the same period of divergence. These distances, however, do not report directly on the rates of evolution.

In order to assess rates of evolution, we measured the amount of time in millions of years for a 1% divergence in amino acid sequence (UEP) (Figure 3) (Dores et al. 1999). While the magnitude of sequence divergence as a function of divergence time appeared to support a molecular clock model, here we find that more recently diverged species appear to be evolving more rapidly, in agreement with previous observations (Pagel et al. 2006; Venditti and Pagel 2010). We then examined two box plots for two different ranges of divergence times (Supplementary Figure 5). As shown in Figure 3, mean UEP values are significantly different ($P < 0.05$) for the pairs 1.3 to 2.2 pg (fish-fish), 4.5 to 6 pg (frog-fish), 4.5 to 35 pg (frog-salamander) and 6 to 35 pg (fish-salamander). Hence, evolutionary rates and genome size are closely associated: species with larger genomes tend to evolve more slowly than species with smaller genomes.

*Discussion*

We investigated the association of genome size with substitution rates in three vertebrate groups using two separate approaches. Substitution rates at neutral sites (dS/My) were assessed between sister lineages for the *rag1* gene and compared to genome size. We also assessed the difference in the frequency of substitutions in non-synonymous sites of the *pomc* gene from species within each group, and then divided the species into different subgroups according to their respective genome sizes. We then determined mean UEP values, which reflect the amount of time in Myr for a one percent divergence in amino acid sequence.

Both approaches-- lineage specific and lineage non-specific-- revealed a clear association of substitution rate with genome size: larger genomes have lower rates of diversification in the *rag1* gene and the POMC amino acid sequence. These observations support the Hinegardner and Rosen hypothesis that fish with large genomes are evolving more slowly than fish with smaller genomes; and suggest that genome size and mutation/substitution rates are co-evolving in vertebrates. Genome size increases in a passive manner with the mutation rate, and evolves according to a number of different processes including deletions, amplifications and the proliferation of transposable elements (TE) (Lynch and Conery 2003; Oliver et al. 2007; Sun et al. 2012). What, however, are the molecular mechanisms that might explain the variation in substitution rates found here and their corresponding associations with genome size?

The hypothesis that genome size co-evolves with and can have a negative influence on mutation/substitution rates appears to conflict with the long established fact that larger genomes are more prone to mutations induced by ionizing radiation and other agents (Heddle and Athanasiou 1975). The hypothesis also appears to be inconsistent with the established view that large genomes impose a mutation hazard on the organism; presumably because large genomes are more genetically unstable, and are associated with smaller effective population sizes (Lynch 2011). Recently, however, studies have demonstrated that mutation rates vary significantly across the eukaryotic genome in a manner dependent on DNA replication timing: rates of mutation increase with replication timing in all eukaryotes examined so far (Herrick 2011). Hence mutation/substitution rates are highly heterogeneous and compartmentalized both spatially and temporally in the eukaryotic cell.

The variation in mutation rates within the genomes of higher eukaryotes appears to coincide with an increase in both TLS and NHEJ activities toward the end of the S phase of the cell cycle (Mao et al. 2008b; Diamant et al. 2012). In yeast, knocking out TLS abolishes the association between mutation rates and DNA replication timing (Lang and Murray 2011), while abrogating RNR activity suppresses the elevated mutation rates associated with TLS (Lis et al. 2008). RNR and dNTP pool sizes also play an important role in determining which repair pathway (HR or NHEJ) is used during the cell cycles of higher eukaryotes (Burkhalter et al. 2009). In late S/G2 phases of the cell cycle, proteasome-mediated degradation of RNR results in declining dNTP pool sizes (Herrick 2010), and hence a concomitant increase in the activity of mutagenic NHEJ. Together, these observations support the proposal that cell cycle-dependent fluctuations in dNTP pools differentially contribute to mutation rates in late replicating DNA when error-prone TLS and NHEJ acivities rise.

As genome size expanded during evolution, different species are believed to have relied increasingly on NHEJ compared to error-free homologous recombination (HR). NHEJ and related repair pathways in the human germline, for example, account for an important fraction of copy number variants (CNV) (Gu et al. 2008; Conrad et al. 2010), suggesting that these error-prone pathways underlie many genomic alterations that either improve or diminish genetic fitness during the course of evolution. In contrast, only a negligible fraction of germline DNA is repaired by NHEJ in the worm *Caenorhabditis elegans* (C-value: 0.1 pg) (Lemmens and Tijsterman 2011). Consistent with an increase in NHEJ activity as genome size expanded, the average number of introns per gene also increases as a function of the fraction of the genome repaired by NHEJ (Farlow et al. 2011). NHEJ is also involved in transposition (Suzuki et al. 2009), suggesting it directly participates in the TE driven genome expansions that occurred, for example, in salamanders (Sun et al. 2012). These observations support the proposal that NHEJ and related repair systems participate in the underlying mechanisms that drive genome size evolution in the vertebrate germline.

How does the cell compartmentalize mutation rates within the genome during its duplication? In eukaryotes, the intra-S phase checkpoint pathway (ATR-Chk1) mediates between DNA replication and DNA repair and slows or stops replication when DNA damage occurs (Despras et al. 2010). During normal S phase, the checkpoint system also participates in mediating the rate at which DNA replication origins fire, and hence influences the rate of progression through S phase (Herrick 2010). Error-free HR repair depends directly on the checkpoint effector Chk1 (Sørensen et al. 2005), and increases steadily with Chk1 activity at the beginning of S phase until HR peaks at mid-S phase (Karanam 2012). As Chk1-dependent HR activity declines during late-S phase, error prone NHEJ and TLS activities steadily rise and persist through mitosis until the beginning of the next S-phase (Mao et al. 2008b). Thus, the intra-S phase checkpoint plays an important role in partitioning mutation rates in the genome between early and late replicating DNA.

The relationship between checkpoint activity and generation time remains to be investigated, but differences in generation time might also reflect a variation in checkpoint activity and differential chromatin content between different species (relative amounts of HC and EC). In species with relatively small genomes but long generation intervals and low diversification rates, such as those found among some of the cartilaginous fish (Figure 2B), a relatively proficient or strong checkpoint function in conjunction with other chromatin regulators might account for the low rates of diversification in those lineages. This proposal, however, remains to be demonstrated.

Indeed, the frequently reported lower incidence of cancer in sharks and salamanders compared to other vertebrates might reflect longer generation times in these species and/or more proficient checkpoint/repair systems: large genomes, or organisms with slower S phases (longer GT), provide error free HR and other repair systems additional time to repair errors in gene rich early replicating DNA (Wintersberger 2000; Mao et al. 2008a; Herrick 2011). This proposal is consistent with the observation in cancer cells that many of the oncogenes that drive tumourigenesis are located in early replicating DNA, and experience lower rates of non-selectable, or neutral, mutation (De and Michor 2011; Woo and Li 2012). Although sharks and salamanders do get cancer, the question remains whether or not the lower incidence of cancer compared to other species is due to correspondingly lower mutation rates in early replicating oncogenes. With the advent of fully annotated and sequenced genomes from higher eukaryotes, addressing these and other questions has now become feasible.

*Methods*

**POMC analysis.** The POMC gene consists of a reading frame of 160 to 260 amino acids, and contains four to five highly conserved domains (γ-MSH, α-MSH/ACTH, β-MSH and β-ENDO) and three domains that are more conserved within each group (Dores and Baron 2011). Amino acid sequences for the POMC gene for 73 species and two outliers (mouse and lamprey) were obtained in the UniProt database (The UniProt Consortium, www.uniprot.org). Sequences were aligned in ClustalW and Mega5 and manually verified by comparison with the previously published alignments (Supplementary Figure 6) (Dores and Baron 2011). Evolutionary divergence between each pair of sequences within each group was estimated by pairwise deletion analysis, and was used to construct a tree for each lineage (Supplementary Figure 7). The species within each group were then subdivided into different subgroups according to their respective genome sizes. The distances for pairs of species within each subgroup were then plotted as a function of the subgroup's average genome size (Supplementary Figure 3). In order to assess rates of evolution, we measured the amount of time in millions of years for a 1% divergence in amino acid sequence (UEP) (Dores et al. 1999).

**Rag1 analysis.** We obtained the *rag1* gene sequences from popsets on the PubMed database (See Table 2 for reference numbers and citations of original publications). One sequence for each species was used. When more than one sequence was present the longer sequence was selected. The sequences were aligned by codon in Mega5 with manual input to verify the presence of stop codons and the correspondence with the amino acid sequence of the Rag1 protein from the same popset. These alignments were used to create a phylogenetic tree by the Neighbor-Joining method (Saitou and Nei 1987). The confidence probability (multiplied by 100) that the interior branch length is greater than 0, as estimated using the bootstrap test (1000 replicates) is shown next to the branches (Supplementary Figure 8) (Rzhetsky and Nei 1992; Dopazo 1994). The evolutionary distances were computed using the number of differences method (Nei and Kumar 2000) and are in the units of the number of base differences per sequence (Supplementary Table 3).

We verified that the lineages in the tree were consistent with the previously published results. The divergence times were obtained either in Time Tree (Hedges et al. 2006)(www.timetree.org) or from published data (see references in Supplementary Table 2). For the lineages for whom we found divergence times we measured the number of synonymous substitutions per synonymous site (dS) and the number of nonsynonymous substitutions per nonsynonymous site (dN) from between sequences using the Nei-Gojobori model (Nei and Gojobori 1986). All ambiguous positions were removed for each sequence pair. Evolutionary analyses were conducted in MEGA5 (Tamura et al. 2011) (Supplementary Table 1).

The species pairs for which both dS and divergence times are available were used to calculate dS/Mya. The C-values for the different species were obtained from the Animal Genome Size Database (Gregory, T.R. (2013). Animal Genome Size Database. http://www.genomesize.com). The statistics of the dS/Mya analysis are shown in Supplementary Table 1.

*Acknowledgements*

Elisa Brambilla for MatLab code for file structure conversion.

*Figure legends*

**Figure 1. Distribution of C-values in fish, frogs and salamanders.** A) Frogs (anura) exhibit a 14 fold range in genome size between one and fourteen pictograms (pg) in a Gaussian distribution. B) and D) Both cartilaginous and ray finned fish exhibit a log normal distribution of C-values. Ray-finned fish exhibit a significantly narrower range of genome size of 0.4 to 5 pg. C) Salamanders (urodela) exhibit a more complex distribution and display two clear peaks. The peaks are centered at 27 and 38 pg correspond to distinct salamander lineages. The C-values in each peak fit a Gaussian distribution (See Supplementary Table 1 for the statistics of the distributions).

**Figure 2.** A) Evolutionary rate (ds/Myr) for frogs, toads and salamanders. Genome size was obtained from the Animal Genome Size Database. Genome size average and median values are shown in Supplementary Table 1. Genetic distances were obtained from aligned sequences (see Supplementary Figure 8 for phylogenetic tree) and divergence times ascertained from the literature (Supplementary Table 2). Although frogs and toads (3 and 7 pg respectively) presented substantially greater genetic diversity, they diverged less recently. Salamanders, in contrast diverged more recently from each other. Although frogs (3 pg) and plethodontidea (20-40 pg) appear to have similar evolutionary rates, hydromantes with a substantially larger genome (42-76 pg) are evolving at a significantly lower rate. B) Similar analyses were performed on tetraodontiformes (T) (average C-value: 0.5 pg), cypriniformes (1.5 pg), skates and rays (SR) (4 pg) and lantern sharks (S) (12 pg). A clear difference in evolutionary rates associated with genome size is apparent. Note that skates, rays and sharks all have exceptionally low and similar evolutionary rates. Inset: log transformed data indicates a power law relationship between evolutionary rates and genome size across these samples. The exponent is -0.39, suggesting significantly different modes of evolution in fish with small genomes compared to fish with larger genomes.

**Figure 3. Box plot of evolutionary rates (EUP) versus genome size.** Units of Evolutionary Period, which reflect the rate of evolution in the POMC gene, increase with genome size, indicating that lineages associated with larger genomes have correspondingly slower rates of evolution. The box corresponds to the middle 50% of the data, and the whiskers to 80%; the small square to the mean and the line to the median. Mean UEP values are significantly different ($P < 0.05$) for the pairs 1.3 to 2.2 pg (fish-fish), 4.5 to 6 pg (frog-fish), 4.5 to 35 pg (frog-salamander) and 6 to 35 pg (fish-salamander). Recently diverged salamanders (far right) appear to be evolving faster than other salamanders that diverged earlier; recently diverged fish (2.2 fish, salmonidae) are also evolving faster than other fish that diverged earlier, see (Pagel et al. 2006). Note, however, that salamanders are evolving more slowly than salmonidae despite the lineage having diverged at about the same time. A clear trend of slower rates of evolution in older lineages is also apparent in each group. Lineage specific effects on evolutionary rates are also apparent independently of genome size: salmonidae (C-value 2.2 pg) are evolving at a faster rate than other fish lineages. Likewise, cartilaginous fish (Heterodontus francisci) and different members of the Actinopergyii class (C-value between 5 and 7 pg) are evolving much more slowly than the other lineages, suggesting that Acipensers qualify as living fossils.

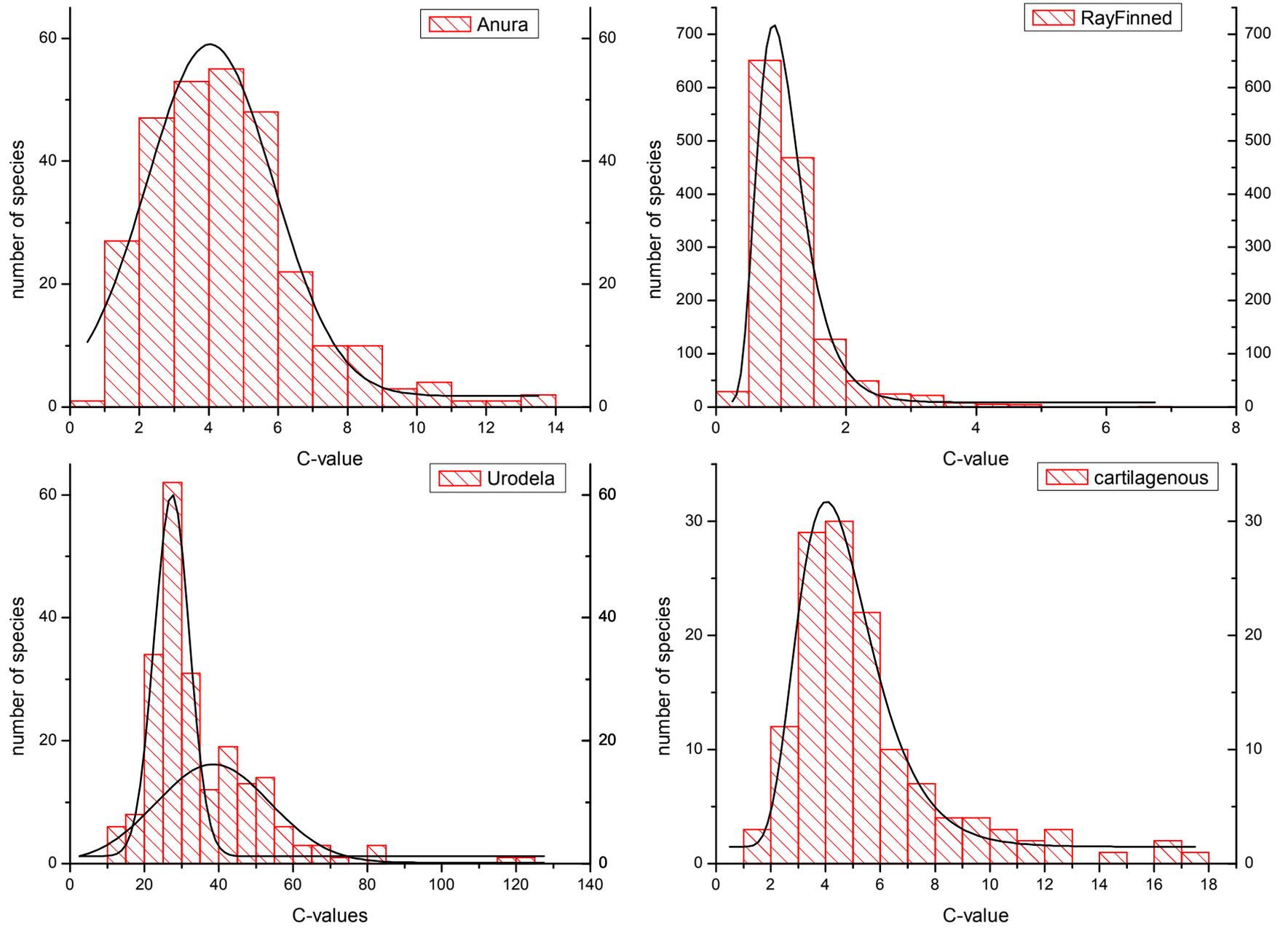

Figure 1

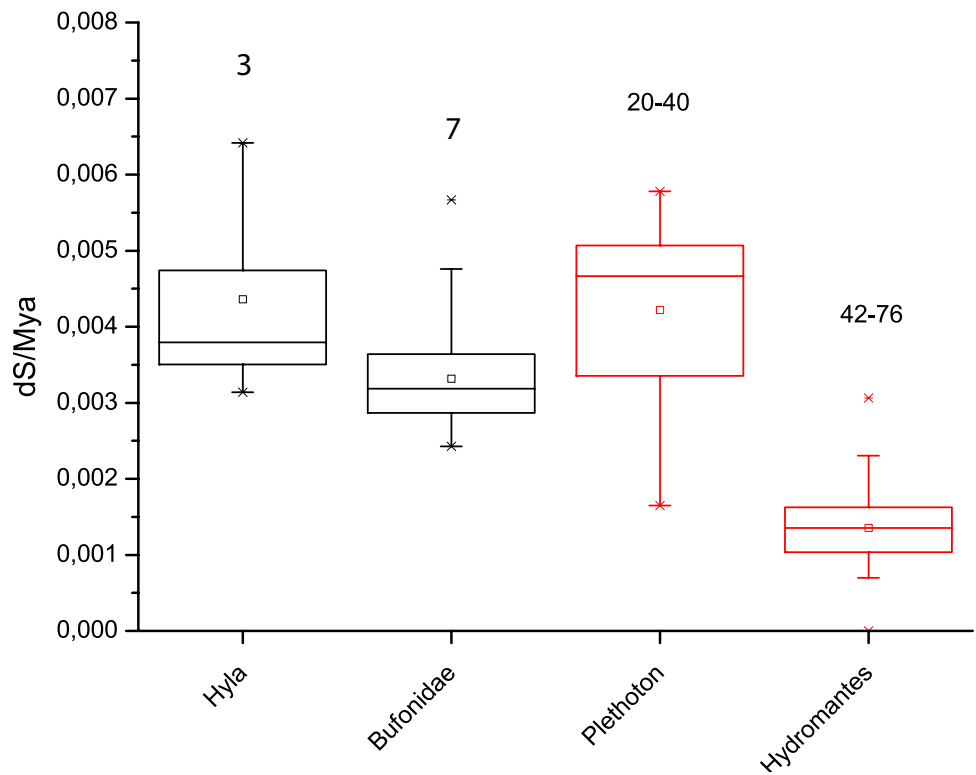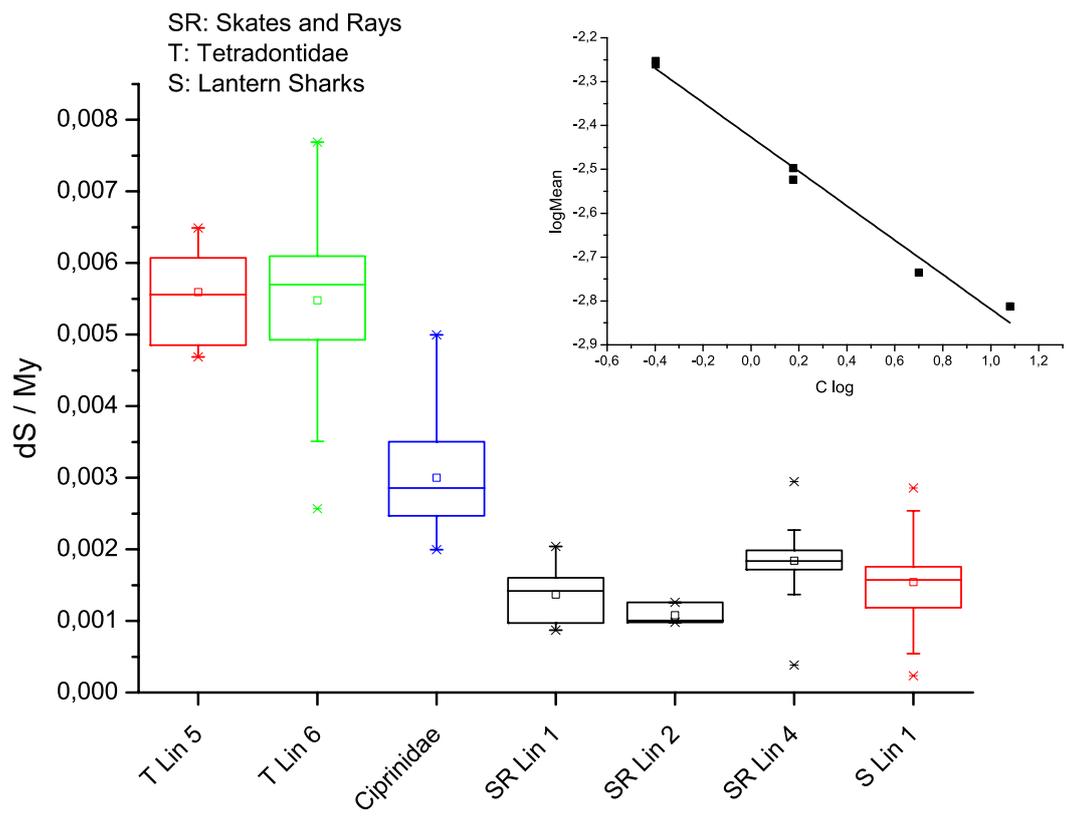

Figure 2

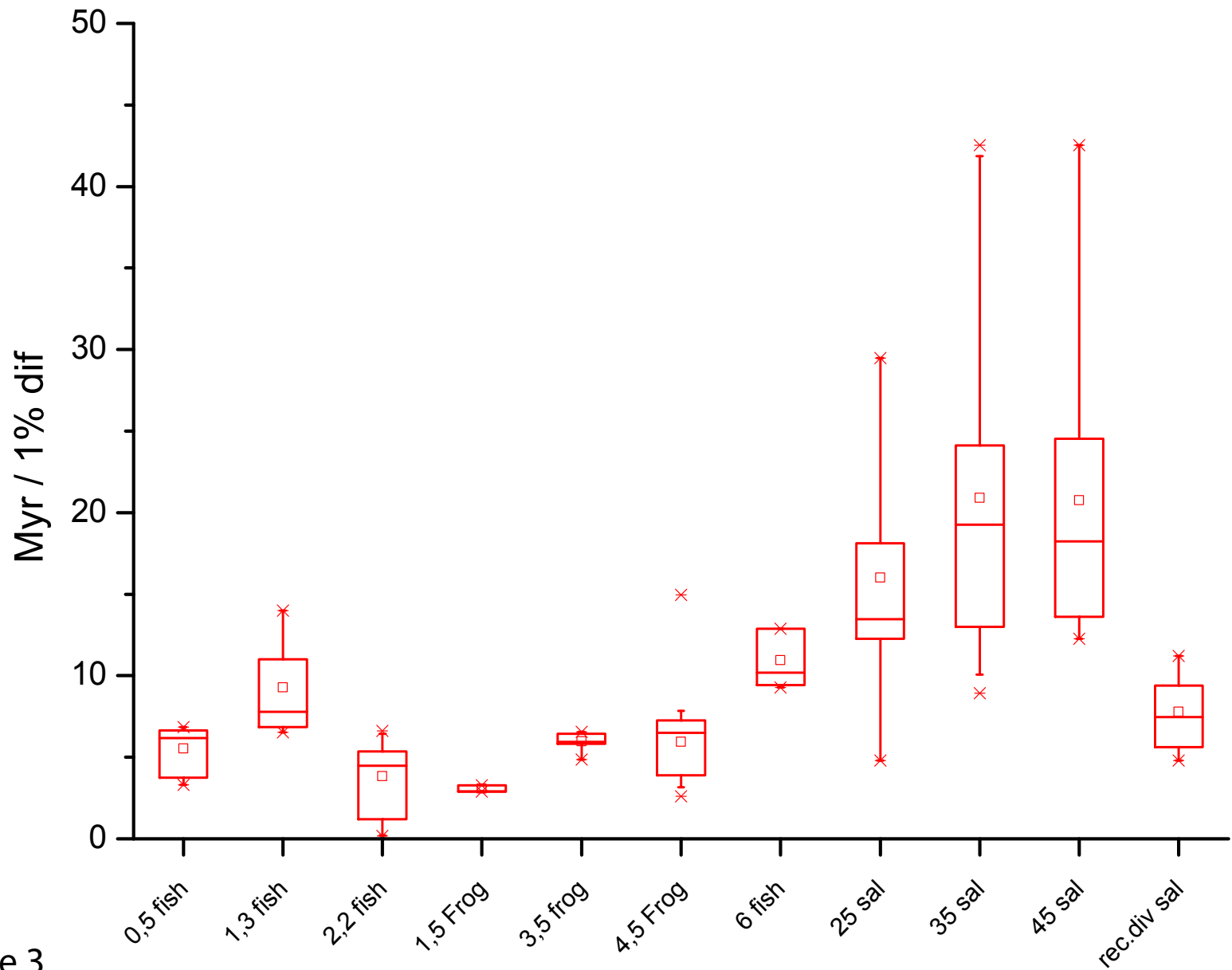

Figure 3